\documentclass[12pt,preprint]{aastex}
\usepackage{amsmath, amsthm, amssymb}

\newcommand{\bepposax}{{\it BEPPOSAX}} \newcommand{\xmm}{{\it
    XMM-Newton}} \newcommand{\epic}{{\it EPIC}} \newcommand{\rgs}{{\it
    RGS}}  \newcommand{\pn}{{\it pn}}
 \newcommand{\chandra}{{\it Chandra}}

  \def\ltsima{$\; \buildrel  <  \over \sim
  \;$}    \def\simlt{\lower.5ex\hbox{\ltsima}}    
\def\gtsima{$\;      \buildrel      >      \over      \sim      \;$}
\def\simgt{\lower.5ex\hbox{\gtsima}}      

\begin{document} 
  
\title{XMM-Newton view of the Multi-Phase Warm Absorber in Seyfert 1
  Galaxy NGC985 \thanks{Partially based on observations obtained with
    XMM-Newton, an ESA science mission with instruments and
    contributions directly funded by ESA Member States and NASA.}}
\author{Yair Krongold\altaffilmark{1}, Elena
  Jim\'enez-Bail\'on\altaffilmark{1}, Maria
  Santos-Lleo\altaffilmark{2},  Fabrizio Nicastro\altaffilmark{3,4},
  Martin Elvis\altaffilmark{4}, Nancy Brickhouse\altaffilmark{4},
  Mercedes Andrade-Velazquez\altaffilmark{1}, Luc
  Binette\altaffilmark{1,5} \& Smita Mathur\altaffilmark{6} }

\altaffiltext{1}{Instituto de Astronomia, Universidad Nacional
  Autonoma de Mexico, Apartado Postal 70-264, 04510 Mexico DF,
  Mexico.}  \altaffiltext{2}{ESA XMM-Newton Science Operations Center,
  ESAC, PO Box 78, E-28691, Madrid, Spain}
\altaffiltext{3}{Osservatorio Astronomico di Roma, INAF, Italy.}
\altaffiltext{4}{Harvard-Smithsonian Center for Astrophysics, 60
  Garden  Street, Cambridge MA 02138, USA.}   
\altaffiltext{5}{Departement de Physique, de Genie Physique et d'Optique, Universite Laval, Quebec, QC G1K 7P4, Canada}
\altaffiltext{6}{Ohio State University, 140 West 18th Avenue, Columbus, OH 43210, USA.}

\begin{abstract}
We present an analysis of a new \xmm\ observation of the Seyfert 1
Galaxy NGC~985. The \epic\ spectra present strong residuals to a
single power-law model, indicating the presence of ionized absorbing
gas and a soft excess. A broad-band fit to the \epic\ and
\rgs\ spectra shows that the continuum can be well fit with a
power-law ($\Gamma\approx 1.57$) and a blackbody component
($kT\approx0.09$ keV). The \rgs\ can be modeled either with two or
three absorption components. In the two absorber model the
low-ionization one, with
$\log$U$\approx.05$ and $\log$N$_H\approx21.08$ accounts for the
presence of the Fe M-shell
unresolved transition array (Fe VII-XIII), and the high ionization
component ($\log$U$\approx1.31$ and $\log$N$_H\approx21.99$) is
required by the presence of several Fe L-shell transitions. 
The data suggest the presence of a third ionized
component with higher ionization, so that the Fe L-shell absorption
features are produced by two different components (one producing
absorption by Fe XVII-XX, and the other absorption by
Fe~XX-XXII). However, the presence of the third absorbing component
cannot be detected by means of an isolated absorption line in a
significant way, so we consider this detection only as
tentative. Interestingly, all ionization components have similar
kinematics (with outflow velocities $\sim 280$ km s$^{-1}$).  In
addition, whether two or three absorbers are considered, the
components appear to be in pressure balance within $1\sigma$. These
results give further support to the idea that warm absorbers in AGN
consist of a two or three-phase medium. We note that, while in the
model with only two absorbers one of them (the high ionization
component) lies on an unstable branch of the thermal equilibrium
curve, in the model with three absorbers all of the components lie on
stable branches of the curve. This gives further plausibility to a
multi-phase absorber.     

\end{abstract}


\section{Introduction \label{par:intro}}

Highly ionized (or `warm', WA) absorbers are observed in about half of
X-ray spectra of broad line active galactic nuclei (AGN), both Seyfert
1s (e.g.  Halpern 1984; Reynolds et al. 1997, George et al. 1998) and
quasars (Piconcelli et al. 2005). The absorption lines of these
systems are blueshifted with respect to the optical emission lines
(with velocities of the order of a few hundreds to $\sim 2000$ km
s$^{-1}$), implying outflowing winds. Their frequency, combined with
evidence for transverse flows (Elvis 2000; Crenshaw et
al. 2003; Arav 2004) suggest that WAs are actually ubiquitous in AGN, but become
directly visible in absorption only when our line of sight crosses the
outflowing  material.  Ionized absorption is also observed in the UV
band.  However, the exact relation between the X-ray absorbers and
Narrow Absorption Line (NAL) UV systems is still uncertain. These
absorbers must be related, as exactly the same AGNs show both (Mathur
et al. 1995), and in many cases, with similar outflow velocities and
similar ionization state.

Several studies have shown that the absorbers can be described in a
simple way. It has been shown that X-ray ionized absorbers can be well
modeled by   including only two or three absorbing components
(Krongold et al. 2003, 2005a;  Netzer et al. 2003; Blustin et
al. 2003;  Steenbrugge et al. 2005). It has
also been found that these components  appear to be in pressure
balance with each other (Krongold et al. 2003,2005a,2007; Netzer et
al. 2003). This has been interpreted as evidence of a multi-phase
medium for the structure of the absorber. Alternatively, it has also
been suggested that we could be looking at a single radially
stratified medium with constant pressure (Rozanska et al. 2006). Other
studies suggest that the warm absorber could consist of a continuous
radial flow of ionization structures (Behar et al. 2003; Ogle et
al. 2004; Steenbrugge et al. 2005), though variability studies seem to
rule out this possibility (Krongold et al. 2005b, 2007). An intermediate
case consisting of a double peaked continuous distribution of material
has also been reported for Mkn 279 (Constantini et al. 2007), although Fields et
al. (2007) reports that these data is consistent with a multiphase
medium, once metalicity effects are taken into account. The exact
nature of AGN winds still remains a mystery.

The nature of the wind directly impacts our understanding of both the
structure and physics of the active nucleus itself, and the effect of
AGNs on  their larger scale galactic and extragalactic environment.
One of the key questions is where do these absorbers
originate. Several studies have set lower limits on the location of
the ionized outflows, placing them at
parsecs from the central ionizing source (Behar et al. 2003, Netzer et
al. 2003), and suggesting an origin in the putative obscuring torus
(Krolik \& Kriss et al. 2001, Blustin et al. 2005). However, other
studies have found WA much closer in, at sub-pc scales (Pounds et
al. 2003; Kaastra et al. 2004, Krongold et al. 2007), leading to the
suggestion that warm absorbers originate from the accretion disk
(Krongold et al. 2007). The mass loss rate inferred for
these outflows is still uncertain by orders of magnitude and is
strongly debated, as this rate depends critically on the distance of
the outflow to the central source. 

\subsection{The Ionized Absorber in NGC 985}

NGC 985 (Mrk 1048) is a Seyfert 1 galaxy located at redshift
0.04274$\pm 0.00005$ (12814$\pm 15$ km s$^{-1}$; Arribas et al. 1999,
based on stellar absorption features, see Krongold et al. 2005 for a
detailed description of NGC 985). 



An ionized absorber has been observed in the X-ray spectra of this
object; 
first through low resolution data (Brandt et al. 1994 in a ROSAT-PSPC
spectrum of this source; Nicastro et al. 1998, 1999, with an ASCA
observation). High resolution observations carried out with the HETGS
on board the {\em Chandra} X-ray telescope (and performed $\sim$1 year
before the \xmm\ observations discussed here) have confirmed the
presence of the absorber (Krongold et al. 2005 hereafter K05). In this
analysis, it was found that two absorbing components were required to
fit the spectrum of NGC 985. Furthermore, the gas pressure of the two
components was consistent, suggesting that the absorber could be
formed by a two-phase wind. The UV band also shows the presence of
absorption lines produced by ionized gas (Arav 2002). Five components
are observed in outflow, with velocities ranging from 780 to 243 km
s$^{-1}$. A sixth component is observed inflowing at -140  km s$^{-1}$.


In this paper we present the analysis of the ionized absorber in
NGC985 carried out on XMM-Newton X-ray spectra. In \S  \ref{par:data}
we describe the reduction of the data. In \S \ref{par:analysis}, we
present the data analysis, including a detailed model of the ionized
absorber in NGC 985.  In \S \ref{sec:disc} we discuss our results, and
compare them with those obtained with the {\em Chandra} data by
K05. Finally, in \S \ref{sec:conc} we present our conclusions.
  


\section{Observations and Data Reduction  \label{par:data}}
The  \xmm\ (Jansen  et  al.  2001)  observation  of NGC\  985 (Obs-Id.
0150470601) was  performed  on July 17,   2003.  Thin filter and small
window mode were applied for the \epic\ (European Photon Imaging
Camera) \pn\ exposure. The {\it MOS-1}
and {\it MOS-2} observations  were performed with the thick and thin
filter  and the partial window mode  and  prime full window mode,
respectively.  Both \rgs\ (Reflection Grating Spectrometer) cameras
were in spectroscopic mode.   {\it
  Science  Analysis Subsystem}, SAS,  v.7.1.2 (Gabriel et  al. 2004)
and the most updated calibration files available in March 2008 were used
to process the data.  According  to the {\it   epaplot} task, no sign
of pile-up  was  detected  in any of   the  \epic\ observations.
Time intervals for which background flaring degrades the signal to
noise in  the \epic\  observations   were excluded  from  the analysis
following   the    filtering   method  described   in   Piconcelli  et
al. (2004). The exposures after the filtering are 40.3, 47.5 and 57.1
ks for \pn, {\it MOS-1} and {\it MOS -2}, respectively. The exposure
for both \rgs\ is $57.7$ ks.


\section{Spectral Analysis \label{par:analysis}}

We have   performed  a spectral  analysis of    the \epic-\pn\ and
\rgs\ instruments  on board \xmm. The circular  extraction regions of
\epic-\pn\  data were centered on the peak of the X-ray emission and
have a radius of 35\farcs7   (750 pixels).   The  background  regions
were selected to be on the same CCD, but far enough from the source to
prevent contamination. 
\rgs\ spectra were extracted using the SAS task {\it  rgsproc} run
with  default parameters. The \epic-\pn\  data have been studied in
the 0.35-10\  keV  (1-37 \AA) band,  and the \rgs\ in the 0.33-1.77\ keV
(7.0-38.0 \AA) band. In  order to apply  the modified Chi-Squared
minimization technique in the spectral analysis, all \epic\ data were
grouped such that each bin  contains at least 20 counts.  The
\rgs\ spectra were fitted with a 10 channel binning 
(i.e. in bins of size 0.1 \AA)  using Chi-Squared
Gehrels statistics. When grouping the RGS data in this way, each channel has at
least 20 counts, and thus possible bias introduced by the use of Chi-Squared
statistics with low count rate data are avoided (see Wheaton et
al. 1995 for the case of Poisson statistics). However, this bin
size is about twice the RGS FWHM, which leads to
a sensitivity loss to detect weak narrow absorption or emission lines in the
spectrum.\footnote{We note however, that a test on our final model (\S
  \ref{final}), using data with a binning size of 0.05 \AA (similar to the size of the
  RGS FWHM), and C-statistics gives results fully consistent with those found
  in our analysis. This indicates that this sensitivity loss has no effect on
  our conclusions.}

The spectral
analysis were performed using the {\it Sherpa}
package\footnote{http://cxc.harvard.edu/sherpa/} of {\it Ciao 3.3}
(Freeman et al. 2001).  We exploit the PHASE code (Krongold et
al. 2003) to model  in a self consistent way the ionized absorbers
present in NGC\ 985. We have explored only photoionization equilibrium
models and we have assumed solar elemental abundances (Grevesse \& Noels 1993). In all the models, we have taken into account Galactic
absorption by applying a cold absorption component   with an
equivalent   Hydrogen column    frozen  to  the   Galactic   value
(N$_H$=3.0$\times10^{20}$\ cm$^{-2}$, Stark et al. 1992).

\subsection{Analysis of the \epic\ data}
The \epic-\pn\  data  was first  analyzed in the hard energy band
(2-10\  keV). Table~\ref{tab:models} shows the values of the continuum
parameters and the goodness  of the fits for several tested models.  A
power law model (model  A) does not  provide an acceptable fit  to the
data, reduced $\chi^2_\nu=1.33$  for 174  d.o.f. 
The most prominent  residuals appear around 6-6.3~keV.  An emission
line improves  the fit (model B) with  a probability  $>99.999$\%
according to an F-test. The energy of the line,  $6.421\pm0.025$ keV
(see  Table~\ref{tab:models}), is consistent with the neutral
FeK$\alpha$ line. The emission line is narrow, $\sigma=0.14\pm{0.03}$
keV. Its  properties are compatible with an origin in low ionization
material, probably located  in the Broad Line Region given the width
of $\sim 6500$ km s$^{-1}$ (this width is too broad to suggest an
origin in the molecular torus). The equivalent width of the line  is
$129^{+25}_{-20}$ eV,  compatible within  the errors with mean values
measured for Type 1 objects (Jim\'enez-Bail\'on et al. 2005). These
are features commonly found in AGNs with WA.  

\begin{figure}
\epsscale{.7} \plotone{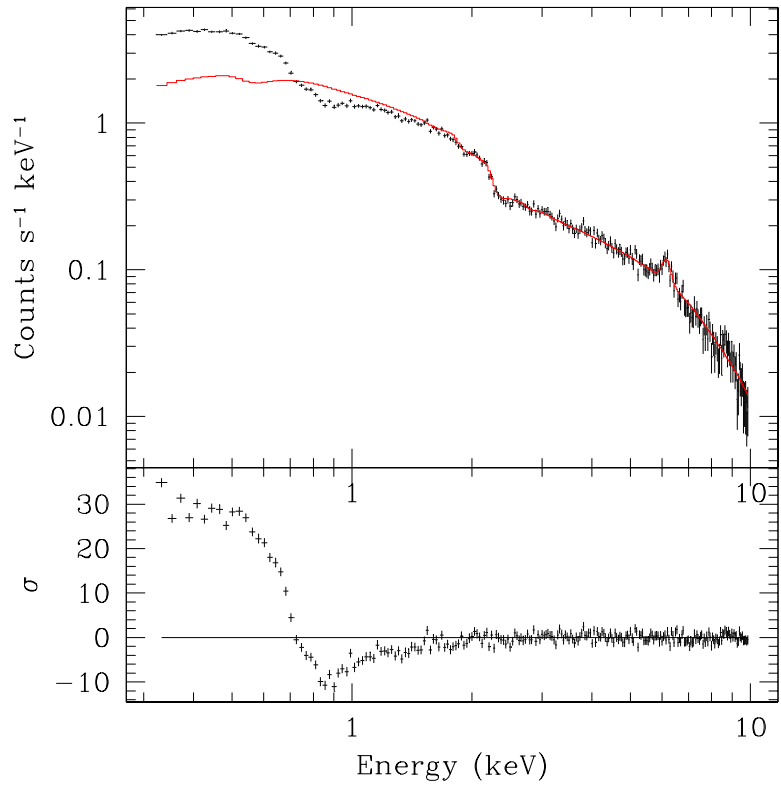} \caption[f1.eps]{ \epic-\pn\
  spectra of NGC~985. Top panel: Model fit in the 2-10 keV energy band
  including a continuum power law plus a Fe K$\alpha$ line, attenuated
  by Galactic absorption. The model (continuous line) has been
  extrapolated to the whole spectral range. Bottom panel: Residuals to
  the model. The presence at low energies of absorption by ionized
  gas, and a soft emission component additional to a simple power law
  is evident from the data.
\label{fig:residuals}}
\end{figure}

Although  model B provides an  adequate fit ($\chi^2_\nu$=0.94 for 171
d.o.f) to  the observed  data  above 2~keV,  when  the lower  energy
band  is considered     the     fit     is strongly     unacceptable.
The lower part of Figure~\ref{fig:residuals} shows the residuals of
the data to model B in the low  energy band. The \pn\ data shows
positive and negative residuals typical of absorbing features  and
also hint the presence of a soft excess below 1~keV.   A cold
absorption component additional to the Galactic one causes no
significant improve to the broad-band fit.

\subsection{Analysis of the RGS data}

Based on the results obtained with the \epic\ data, we have tested for
the presence of the ionized absorber in the \rgs-\xmm\ high resolution
data. The continuum model,
consisting of a power law plus a blackbody (as indicated by the
\epic\ data), was first attenuated by one warm absorbing component modeled with
PHASE (model C). PHASE has 3 free parameters for each absorbing
component: the ionization parameter ($\log~U$), the equivalent
Hydrogen column density $(\log$~N$_H$) and the outflow velocity
(v$_{out}$) of the absorbing gas. A fourth parameter, the
microturbulence velocity of the material ($v_{Turb}$, which is added in quadrature
to the thermal velocity to compute the total Doppler broadening:
$v_{DOP}^2=v_{Therm}^2+v_{Turb}^2$) is not easily constrained
with current grating spectra and, in all the models used here, was set
equal to 300 km s$^{-1}$ (see Krongold et al. 2003 for details)\footnote{We
note, however, that the value of this parameter has little effect on
our results. For instance, a test using a value of 100 km s$^{-1}$ for
this velocity gives values for the free parameters fully consistent
(within  1$\sigma$) with those obtained in our models.}. 

The spectral energy distribution (SED) used to produce the PHASE
models was the same used by K05 (see their Fig. 3a and \S 2.1.2),
based on multi-frequency data of NGC~985 (mostly obtained from the
NASA Extragalactic Database, NED)\footnote{During the \xmm\ observation the
  X-ray spectral properties of the source were strikingly similar to
  those of the 2001 \chandra\ observation (see \S \ref{sec:disc}). It
  is well known that the absorbing properties of the gas in the X-ray
  region weakly depend on the shape of the SED below 0.1 keV (Netzer
  1996; Steenbrugge et al. 2003). This justifies using the same
  SED  used by K05, which provides a direct comparison between models,
  rather than constraining the UV shape of the SED using the optical
  monitor (OM) data of the \xmm\ observation.}.  This SED has a UV
break beyond 1200\AA\ (at 912\AA), and the UV and X-ray continua meet
at around 0.1 keV.  This shape is consistent with recent results by
Haro-Corzo et al. (2006), who studied in detail the shape of the SED
from the far UV to the X-rays.

Model C further included 4 narrow emission lines, required by
the data, and compatible with being the OVII triplet and the OVIII
K$_\alpha$ line.  The best fit values of the parameters of the ionized
absorber for model C (and for all models tested) are shown in
Table~\ref{tab:ioniz}. The continuum parameters and the goodness of
all the models tested are reported in Table \ref{tab:models}.  This
fit, though acceptable ($\chi^2_\nu$=1.16 for 410 d.o.f.), still
leaves the presence of residuals due to absorption. This is consistent
with the findings by K05, who reported that two absorption components
were required to fit the data.

To test the presence of this second absorber, we further added to
model C a second PHASE absorbing component. Model D ($\chi^2_\nu$=1.12
for 407 d.o.f.) significantly changed the fit with respect to a
single absorber.  An F-test gives a probability $\approx99.9$\% for the
presence of the second absorbing component.

The low ionization absorbing component found in the \xmm\ spectra of
NGC 985 (hereafter low-ionization phase or LIP) can be clearly
identified by its characteristic absorption by the Fe M-shell (Fe-VII-XIII)
unresolved transition array (hereafter UTA, see Behar et al. 2001), and by the
presence of the O~VII resonant line at $\sim$18.6 \AA. This component should also produce
absorption by charge states like O~VI, N~VI, and C~VI. Lines by these ions are
not clearly detected in the spectrum. However, our model does include them,
and the data is consistent with their presence (see also \S \ref{emlin}). 
The high ionization absorbing component (high-ionization phase or HIP)
is detected by several Fe L-shell charge
states, in particular by Fe XVII-XXII.


Few sources show the presence of a very high ionization absorbing
component (e.g. NGC 3783, Netzer et al. 2003; NGC 5548, Steenbrugge et
al. 2005). K05 suggested the presence of such a component in NGC~985,
but reported that it could not be constrained with the {\em Chandra }
data. 
Thus, we further added a third ionized absorber to our fits (model E),
to test the possible presence of this component.
Model E improves by $\Delta \chi^2$=18 over model D,
for a difference of 3 d.o.f (the ionization parameter, column density,
and outflow velocity of the third absorber). According to an F-test
this difference is significant (significance  $\sim99.9\%$).

The ionization parameter and column density of the LIP remain
consistent within $1\sigma$ between this fit and model D. However, the
ionization parameter of the HIP decreases by a factor $\sim 0.60$, and
the column density of this component decreases by a factor  $\sim
0.42$. The leading charge states of the HIP in model E are then Fe
XVII-XX. In turn, the leading charge states
of the 3rd, hottest component (hereafter super-hot ionization phase,
or SHIP) are  Fe XX-XXII (the
leading charge states of the LIP are the same ones listed above, as
the ionization state of this component did not vary from model D to
model E).

\subsection{A Global fit to the Spectra \label{final}}
\begin{figure}
\figurenum{2}
\plotone{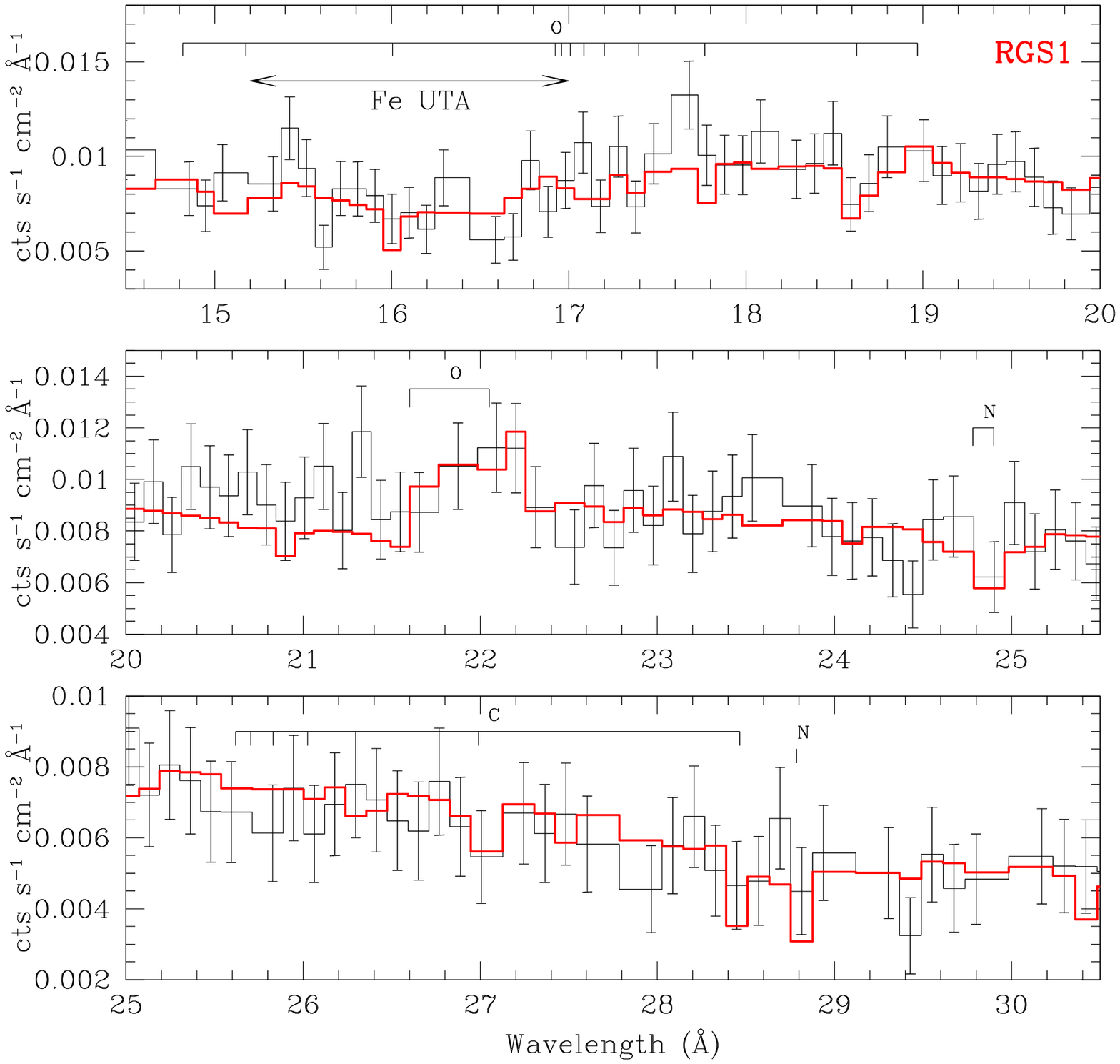} \caption[f2.eps]{Three-phase absorber
  model plotted against the \rgs1\ spectra of NGC~985. The data is
  presented at the rest frame wavelength of the ionizing absorber.
  The model (Model F) was produced by fitting simultaneously the \rgs\ and
  \epic\ spectra.    
\label{fig:RGS1}}
\end{figure}

\begin{figure}
\figurenum{3}
\plotone{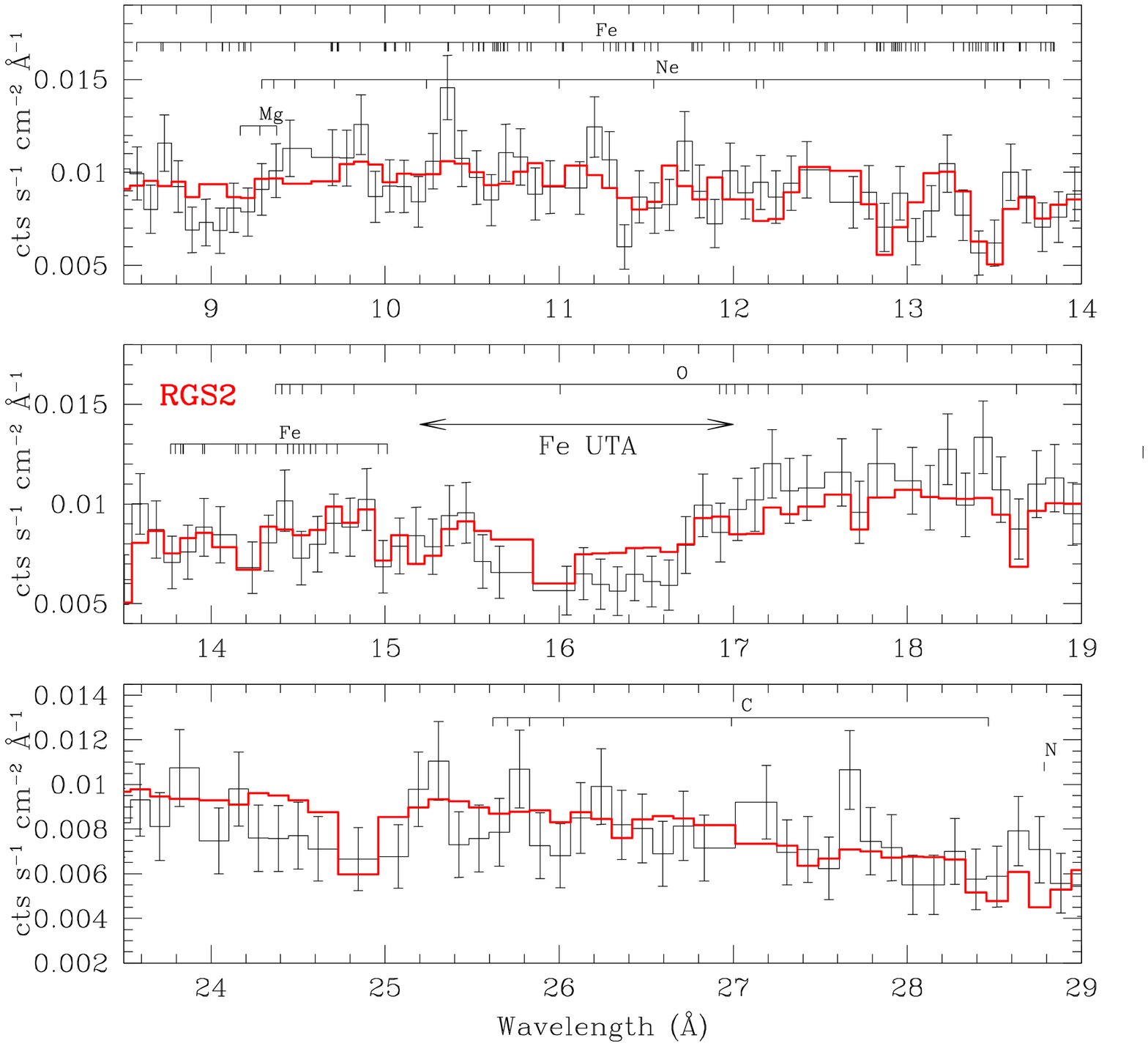} 
\caption[f3.eps]{Three-phase absorber
  model plotted against the \rgs2\ spectra of NGC~985. The data is
  presented at the rest frame wavelength of the ionizing absorber.
  The model (Model F) was produced by fitting simultaneously the \rgs\ and
  \epic\ spectra.  
\label{fig:RGS2}}
\end{figure}

\begin{figure}
\figurenum{4}
\plotone{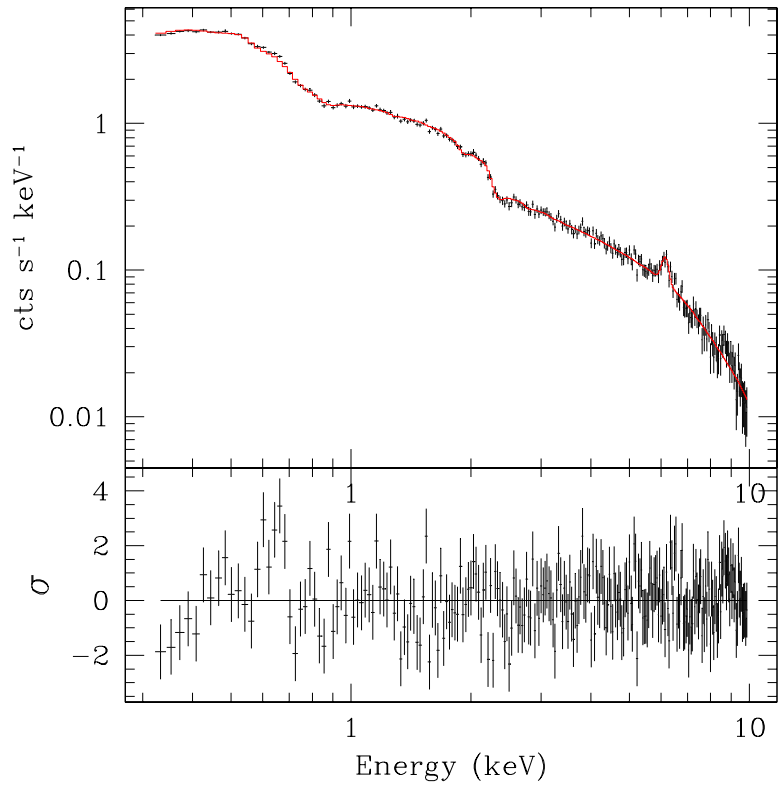} \caption[f4.eps]{Model and residuals of a
  three-phase absorber model (Model F) over the \epic\ data of
  NGC~985. The fit was carried out simultaneously for the \rgs\ and
  the \epic\ data.
\label{fig:EPIC}}
\end{figure}

Finally, we analyzed the \rgs\ and \epic\ data simultaneously  over the whole
energy band. Based on  the results obtained for the independent
analysis performed for \epic\ and \rgs, we fit the spectra (model F)
with a  power law and  a black body component absorbed by three phases
of ionized material (LIP, HIP and SHIP), plus the four narrow emission
lines detected in the \rgs, and the Fe line detected with \epic.  We
forced the photon
index of the power law and the temperature of the blackbody to be the
same in all the models for all the instruments, but allowed the
normalizations to vary freely among spectra from different
detectors. The outflow velocities were fixed to the best fitting
values found for the models performed only to the \rgs\ data. The
parameters for this model are
presented in Tables~\ref{tab:models} and ~\ref{tab:ioniz}.  The
parameters of the four narrow emission lines (OVII triplet and OVIII
K$\alpha$) modeled are collated in
Table~\ref{tab:lines}. Figures~\ref{fig:RGS1}, ~\ref{fig:RGS2},
and~\ref{fig:EPIC}
present data and best fit model for the \rgs1, \rgs2, and \epic-\pn,
respectively. 

This fit ($\chi^2_\nu$=1.25 for 648 d.o.f.) produced values for the ionized
absorbers fully consistent with those derived in model E. However, the photon
index among the different fits do vary in a significant way. This is due to the
high energy cut off of the \rgs\ being below 2 keV, which does not
provide a  sufficient baseline to constrain this parameter well.  
It is interesting to note that the continuum parameters for our final
model are consistent with those reported for this object by K05,
indicating no change in the spectral shape of the source during the
two observations.

\subsubsection{Continuum Fits}

While model E is acceptable over the whole spectrum, in the spectral
region between 17 and 21 \AA\  there is a clear excess of flux, of about
10\% in the data with respect to the model\footnote{There is also an
  overprediction of
  the model between 15.8 and 16.8 \AA, in the region where the UTA is located,
  which is likely produced by the use of the abbreviated data of the Fe
  M-shell transitions that produce the UTA (Behar et al. 2001) in our model,
  see Krongold et al. (2003) for details.}. This excess, which is present in both the RGS
and EPIC data, can be modeled (using an empirical
approach) by adding to model E a broad Gaussian component ($\chi^2_\nu$=1.21 for 645
d.o.f.). This broad component, with position at 20.1$\pm0.6$\AA,
and FWHM$\sim$7.7$\pm1.2\times10^4$ km s$^{-1}$, is required by the data with a
confidence level of 99.99\% (according to an F-test). The width, and position
of the line are clearly not consistent with an origin in the Broad Line Region
of the source. On the other hand, considering the effect of the ionized
absorber below 19 \AA, the excess could be produced by a
relativistic O VIII K$\alpha$ emission line (e.g. Branduardi-Raymont et
al. 2001, Ogle et al. 2004). However, the quality of the data
is not good enough to further test this idea.
Whatever the nature of this emission, we note that it has no effect on the
results presented here for the ionized absorber. The parameters obtained for
the absorbing components including this broad feature are fully consistent
with those found in Model E.

\subsubsection{Absorption and Emission Blending \label{emlin}}

From Figures 2 and 3 it can be observed that, in the regions where the
OVII and OVIII emission lines are located, the model predictions
are in good agreement with the data. However, while a visual inspection to
these spectral regions shows that the emission lines are barely present in
the data,
Table 3 indicates that they are detected with high significance. This effect
is the result of a strong filling of the absorption lines by the emission
features. Indeed,
the spectra do not show the presence of absorption lines at $\sim$18.9 \AA,
$\sim$21.6 \AA, and 22.1 \AA, where the OVIII K$\alpha$, OVII K$\alpha$, and
a strong OVI absorption lines are located. Yet, these three lines are black
saturated in our models. In addition, the OVII K$\beta$ absorption line at
18.6 \AA\ is detected in the data, which means that the K$\alpha$ transition
should also be present. This further supports the idea of emission and
absorption line blending.      

\section{Absorber Discussion \label{sec:disc}}

\subsection{A Static Outflow in Photoionization Equilibrium \label{sec:pe}}

The ionized absorber found in NGC~985 shows similar characteristics to
WAs found in other systems, i.e. the presence of 2 or 3 different
ionization components, with increasing column density as the
ionization level increases. We note that the 2 absorber best fit model
(model D) found for the \xmm\ data is strikingly similar to the one
found for the \chandra\ data. Comparing the best fit ionization
parameters for the HIP and the LIP of model D
with those found by K05 in the \chandra\ data, it is observed that
the values are consistent with each other within $1\sigma$. The column
densities are also consistent  within $2\sigma$. We
also note that the ionizing flux level of NGC~985 during the
\chandra\ and \xmm\ observations is very similar (F($0.1-10$keV)
$=1.2\pm0.2\times10^{-11}$ erg s$^{-1}$ cm$^{-2}$ during the first
observation and F($0.1-10$keV)  $=1.46\pm0.21\times10^{-11}$  erg
s$^{-1}$ cm$^{-2}$ during the latter one). In addition, the continuum
also has indistinguishable parameters during the two observations,
with similar photon indexes and blackbody temperatures. This gives us
a justification for the use of the same SED between our models and
those presented by K05.      

The fact that the spectral shape and continuum level of NGC 985 is the
same between the two observations, together with the finding of
consistent ionization parameters in both spectra strongly suggests
that the absorbing components in this object are in photoionization
equilibrium. In turn, the lack of significant changes in the column
density of the components suggests that we are looking at a static
outflow, that has not undergone significant changes in the $\sim$1 yr
elapsed between the \chandra\ and \xmm\ observations, and probably
within the $\sim4$ yr between the \bepposax\ (also analyzed by K05)
and the \xmm\ observations.

\subsection{The 3rd, Hottest, Ionization Absorbing Component} 

\begin{figure}
\figurenum{5}
\plotone{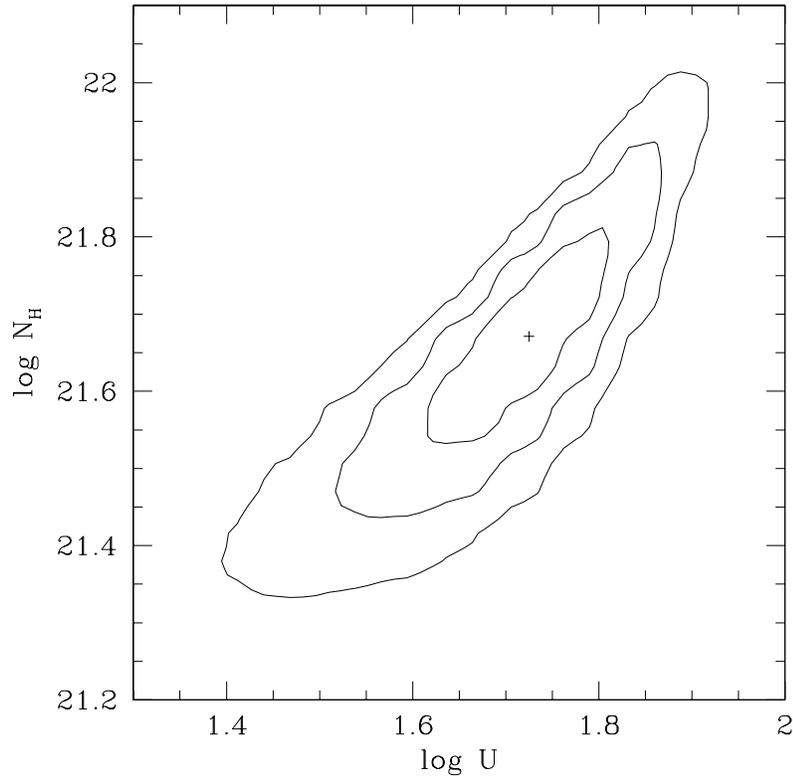} \caption[f5.eps]{Confidence region
  for the ionization parameter vs. the Hydrogen equivalent column
  density for the 3rd absorbing component required by the simultaneous
  fit to the \rgs\ and \epic\ data of NGC~985. The contours represent
  the 1, 2, and 3 $\sigma$ confidence levels. Despite the presence of
  this component cannot be significantly demonstrated by any single
  absorption line, the parameters of the absorber are well constrained
  by the data.
\label{fig:reg_proj}}
\end{figure}

The \xmm\ data shows that the high ionization component can de-blend
into two components (HIP and SHIP). We visually
searched for, but could not identify, a discrete absorption feature
that could show in a significant way the presence of this component in
the \rgs\ spectra. The improvement in the fit when this absorber is
included is produced by
differences smaller than 1$\sigma$ integrated over the whole
spectrum. Despite being unobservable by eye, this component appears to
be well constrained, as shown in Figure \ref{fig:reg_proj}, where 1, 2
and 3$\sigma$ confidence regions for the physical parameters of this
absorber are presented.   

\begin{figure}
\figurenum{6}
\plotone{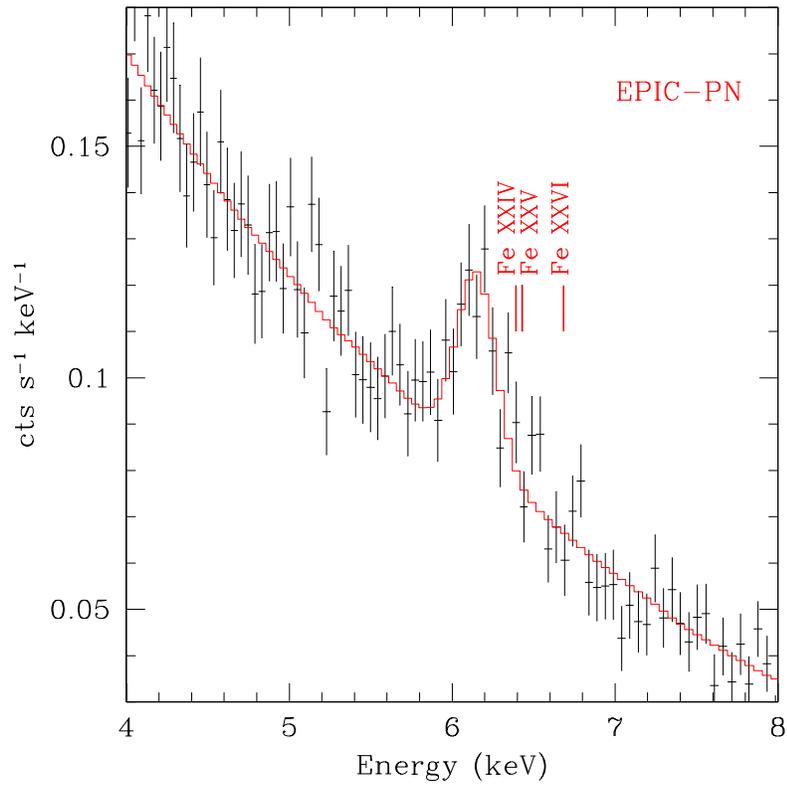} \caption[f6.eps]{Fit to the Fe line region
  of the \epic-\pn\ spectrum of NGC~985. The labels mark the expected
  position of absorption lines by highly ionized Fe (Fe
  XXIV-XXVI). Clearly, no such absorption features are present in the
  spectrum. 
\label{fig:fe_line}}
\end{figure}
\begin{figure}
\figurenum{7}
\plotone{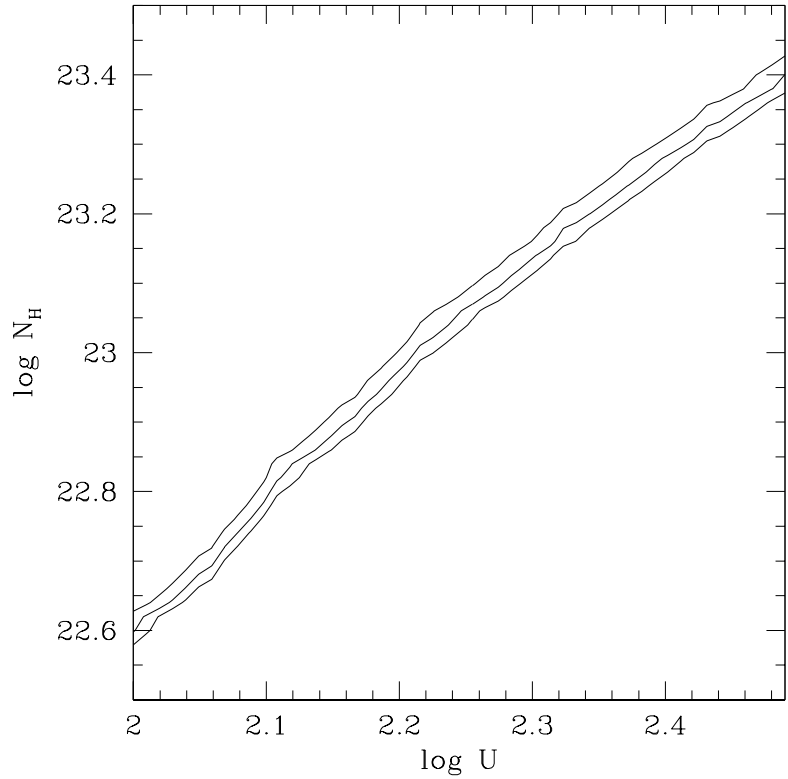} \caption[f7.eps]{Confidence region for the
  ionization parameter vs. the Hydrogen equivalent column density for
  a very high ionization absorbing component. The contours represent
  (from bottom to top)
  the 1, 2, and 3 $\sigma$ confidence levels. Such hot gas would be
  almost transparent to the impinging radiation (as the gas would be
  almost stripped). The data does not allow us, then, to rule out such
  an extreme component. 
\label{fig:reg_proj2}}
\end{figure}

We notice that the limited S/N of the
\chandra\ data of the NGC~985 spectrum was not enough for the
detection of this absorbing component.    
Actually, the SHIP is substantially different from the component hinted by the
\chandra\ data. That component was suggested by small residuals
between 1.6 and 1.8 \AA, where absorption by Fe XXIV to Fe XXVI could
be expected. Instead, the higher S/N of the \xmm\ data in this region
(through the \epic-\pn) allows us to rule out such a component, as no
absorption lines by such charge states are detected (see
Fig. \ref{fig:fe_line}). We set equivalent width $2\sigma$ upper
limits of 5.3, 3.6 and 2.5 m\AA\ to the  presence of absorption by  Fe
XXIV, Fe XXV, and Fe XXVI, respectively. However, the presence of gas
with even higher ionization is still possible.
In Figure \ref{fig:reg_proj2}
we present 1, 2, and $3\sigma$ contour plots for the presence of an
extremely highly ionized absorbing component in NGC~985 (with
ionization even larger than the SHIP and that expected from the
\chandra\ data). The presence of such material cannot be ruled out, as is
formed by almost entirely stripped gas and thus is almost transparent
(for atomic absorption) to the impinging radiation.

\subsection{On the Nature of the Absorber: A Multi-Phase Medium}
\begin{figure}
\figurenum{8}
\plottwo{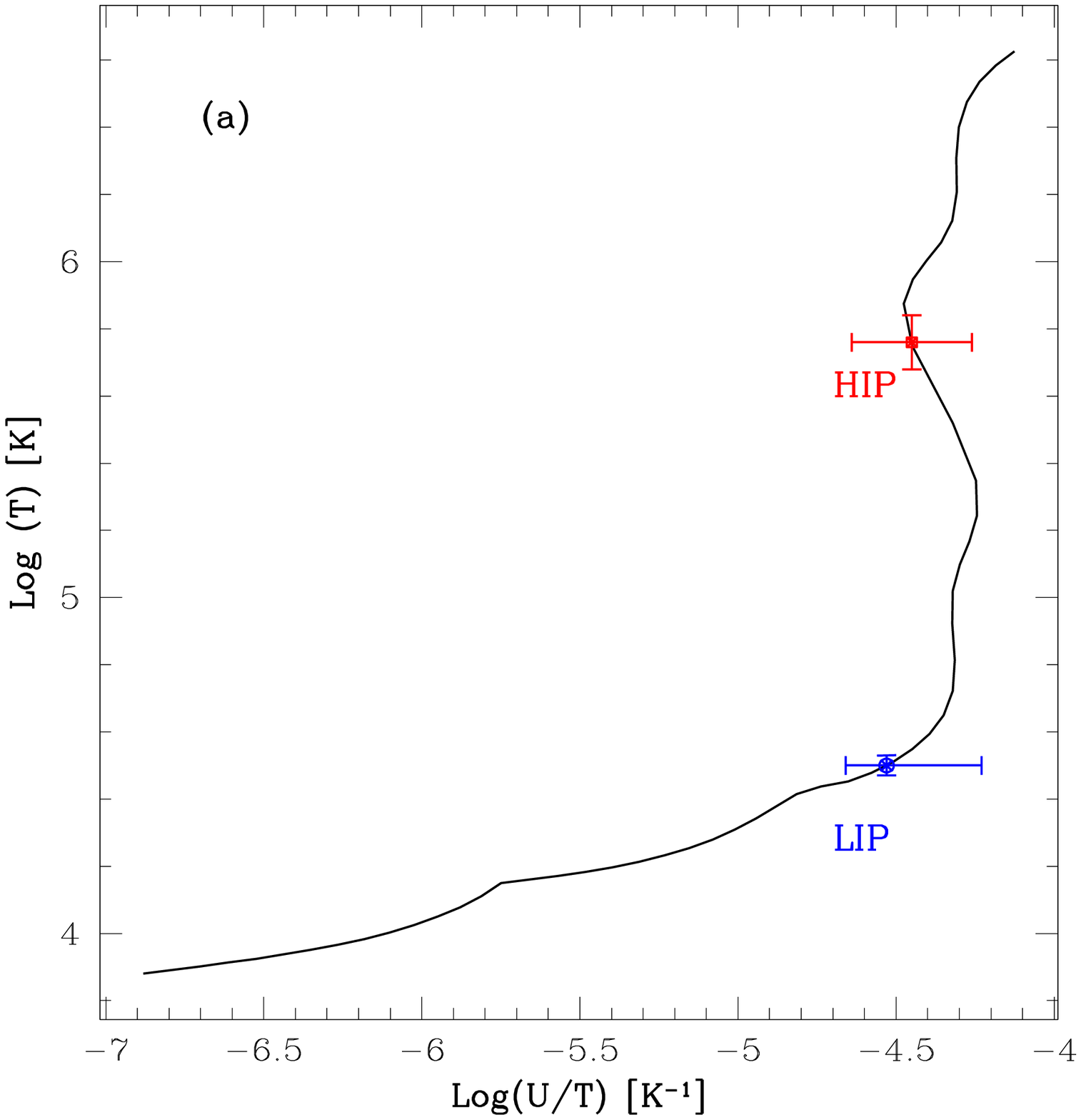}{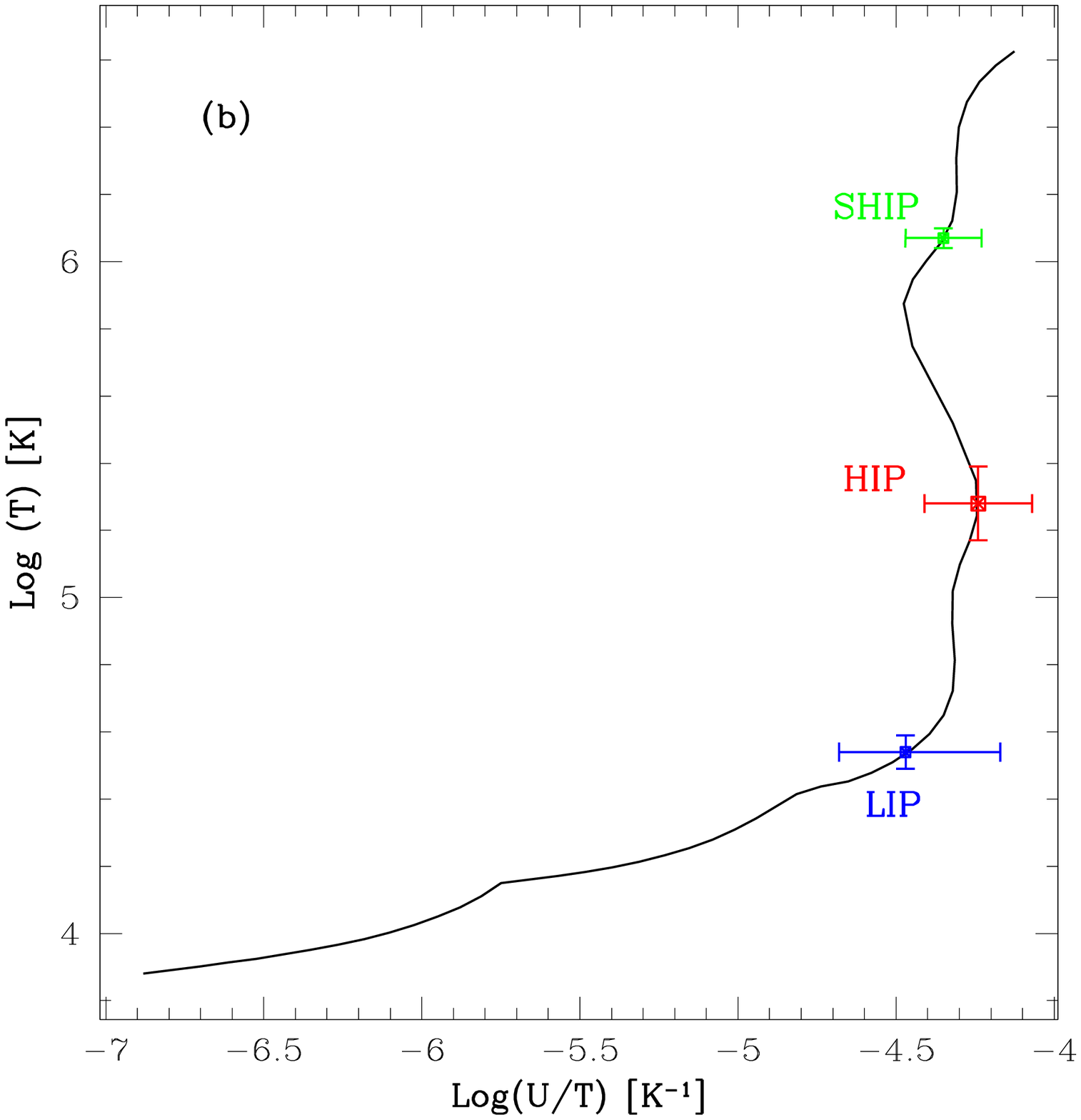}
\caption[f8.eps]{Thermal equilibrium curve of the photoionization
  models used to analyze the data of NGC~985. Panel (a) marks the
  position of the two-phase absorber model (Model E) in the curve. The
  two phases are in pressure equilibrium, but the HIP lies in an
  unstable region of the curve. Panel (b) marks the location of the
  three phases of Model F. The phases are again in pressure balance,
  and they all lie in regions of stability. The large positive error
  bar in the LIP represents the possible effect in the best fit value
  of $\log~U$ induced by the lack of low-temperature dielectronic
  recombination rates for Fe (e.g. Netzer 2004, Kraemer et al. 2004;
  see discussion by K05). \label{fig:xi}}
\end{figure}

K05 reported that the two absorbing components detected in the
\chandra\ data have the same value of the pressure, and thus, if
located at the same distance from the ionizing source, could form a
single medium in pressure equilibrium. One component could then be
confining the other, or a third hotter component could be confining
the two detected ones. This result is confirmed by our analysis of the
\xmm\ data. Figure \ref{fig:xi} presents the thermal photoionization
equilibrium curve (or S-curve, Krolik et al. 1981) for the
SED used in our analysis. The S-curve marks the points of thermal
equilibrium in the T vs. U/T plane,  where T is the photoionization
equilibrium temperature of the gas, and U/T is a quantity inversely
proportional to the pressure of the gas (see Komossa \& Mathur, 2001,
and K05 for a detailed explanation of this curve). As can be observed
in  Figure \ref{fig:xi}a, the best-fit two absorbing components of
model D (HIP and LIP) have very different temperatures, but lie very
close to each other in the U/T axis, implying very similar
pressures. In fact the gas pressures of the two components are
indistinguishable within $1\sigma$.

However, it is interesting to note that the HIP in model D, as well as
the HIP found for both \chandra\ and \bepposax\ spectra of NGC~985 by
K05, lie close to, but out of the region of thermal stability in the
S-curve.  Gas in regions of the curve with negative derivative is
unstable because any isobaric perturbation will be amplified, leading
to net cooling or heating. Thus, the HIP should not be able to survive
in this unstable region of the S-curve for periods of time longer than
the hydrodynamic timescale of the system $t_H$ (i.e. the timescale in
which the medium adapts to mechanical perturbations). K05 points out
that, if the different ionization components form truly a multi-phase
medium, then $t_H$ must be shorter than the photoionization
equilibrium timescale $t_{ph}$ (i.e. the timescale in which the system
can adapt to changes in the ionizing continuum). Otherwise, the
pressure balance between the phases could be easily broken and the
phases would dissolve (see detailed discussion in \S 5.5 of
K05). Since the gas forming the warm absorber of NGC 985 is most
likely in photoionization equilibrium (\S \ref{sec:pe} and K05) this
implies short  $t_{ph}$, and thus short $t_H$. This presents a serious
problem for the existence of the HIP, as the gas that forms it should
go  rapidly to stable regions of the S-curve (the HIP should dissolve
during the 1 year elapsed between the \chandra\ and
\xmm\ observations). One possible (though unlikely) explanation could
be that the ionization parameter (and thus the temperature) of the HIP
has been systematically underestimated in all the analyses of the
X-ray spectra of this source, and that this component really lies on
the intermediate branch of stability. Another explanation is that the
crossing time of the phases through our line of sight is $<< t_H$, and
we always see the phases before they dissolve.

Another explanation that relaxes the above relation between the crossing
time and  $t_H$ arises from the possible presence of the third higher
ionization component (the SHIP) detected only with the larger S/N of the
\xmm\ data\ (Models E and F).  Figure~\ref{fig:xi}b shows the location in the
thermal equilibrium curve of the 3 absorbing components found in Model
F. The pressure of the 3 phases is again indistinguishable within
$1\sigma$, however, in this case all 3 components lie in thermally
stable regions of the 'S-curve.' This 3-phase solution allows the
phases to co-exist for
long periods of time, even if  $t_H$ is small (for instance if the
confined phases are composed of small cloudets, as suggested by K05).
This is consistent with the finding that the 3 absorbing components
have indistinguishable kinematics (the outflow velocity of the 3
phases is consistent with a single value well within $1\sigma$, see
Table \ref{tab:ioniz}). This  indistinguishable kinematics would
further prevent the confined phases to be rapidly destroyed by drag
forces (which depend on the relative velocity among the phases).
These results support the idea that warm (ionized) absorbers in AGN
are formed by a multi-phase medium in pressure equilibrium, with one
hot component confining the others.
In this case, the different phases of the multi-phase medium could
form easily, as the gas tends to go to the stable branches of the
thermal equilibrium curve, while keeping pressure balance between
them.

WA in other Seyfert 1s also show different phases, in pressure balance
(e.g. NGC~3783 Krongold et al. 2003; Netzer et al. 2003; Mrk 279 Fields
et al. 2007, though see Constantini et al. 2007). An obvious
requirement for the phases to form a single medium, is that they
should lie in the same location, i.e. they should be at the same
distance from the ionizing source. We have no means to constrain the
location of the absorbing components in NGC 985 with the present
data. However, we note that, using time-evolving photoionization
arguments,  Krongold et al. (2007) found that the 2 phases in the
ionized absorber in NGC 4051 are consistent with being co-located and
in pressure equilibrium.

\begin{figure}
\figurenum{9}
\plotone{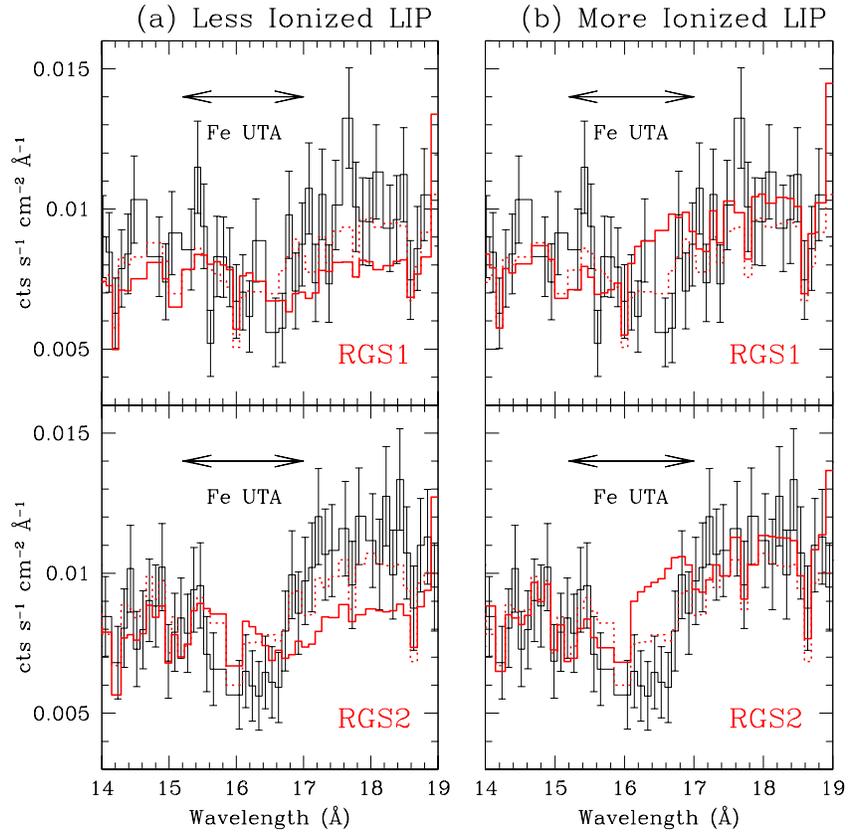}
\caption[f9.eps]{LIP models with different value of the ionization
  parameter. The dotted line shows the best fit model for
  comparison. (a) Less ionized LIP; $\log(U) = -0.4$. (b) More ionized LIP; $\log(U) = 0.3$.  \label{fig:cont}}
\end{figure}

Our analysis does not seem to favor a continuous radial range of
ionization structures. As shown in  Figure \ref{fig:cont}, the data
rules out the presence of significant absorption by Fe I-VI. The
lack of gas with lower ionization state is
further evidenced by the non-detection of O~I-V. We find a 2$\sigma$ upper
limit of 17~m\AA\ for the EW of the O V 22.37 \AA\ transition. This implies a ionic
column density $<7\times10^{15}$ cm$^{-2}$,  a factor $\approx$85 and
$\approx$20 lower than those inferred (from our model) for O~VII and O VI, respectively
(and consistent with the O V column density predicted in our model).
From Figure \ref{fig:cont} it can also be observed that the data is
inconsistent with absorption by Fe XIV-Fe XVI. Thus, at least the LIP
seems to be a genuine discrete component.
The situation for the HIP and SHIP is more complex, as the ionization
parameters of these components are separated only by a factor
$\approx$5. To test whether these two components could be better
reproduced by more components (indicating the possible presence of a
continuous radial flow for the high ionization gas), we modeled the data
adding a fourth ionization component with ionization parameter between those
of the HIP and SHIP. The fit showed that this
component was not required by the data, further suggesting
discrete phases (we set a  2$\sigma$ upper
limit of 1.3$\times10^{20}$ cm$^{-2}$ for the column of this fourth absorber).
These results are
consistent with the findings by K05 who, based on variability, ruled
out a continuous radial flow. 
The idea of a continuous radial range of ionization structures is
recurrent in the literature (e.g. Steenbrugge et al. 2005;
Ogle et al. 2004) as it can explain, with a
single outflow, both the presence of the warm absorber in Seyfert 1
galaxies and the presence of extended emission produced by highly
ionized gas in Seyfert 2 galaxies (Kinkhabwala et al. 2002).  However, a multi-phase medium is
also consistent with the idea of a single outflow, but detected in
emission and absorption at different locations (see Krongold et
al. 2007 for details).

Our results further support a compact absorber.  Combined with strong
evidence from the UV data for perpendicular motion with respect to our
line of sight is several UV absorption systems in other objects
(Mathur et al. 1995;
Crenshaw et al. 2003; Arav et al. 2004), this strongly suggests that
we are looking the flow in a transverse direction, rather than
directly in the direction of the flow. This is consistent with a
bi-conical structure for quasar winds (e.g. Elvis 2000).


\section{Conclusions \label{sec:conc}}

The \xmm\ data of NGC 985 shows the presence of an ionized (warm)
absorber in the X-ray spectra  of this source. The data can be
modeled with two or maybe three absorbing components. The presence of
a third absorber is statistically significant in the models, however we
could not identify  a discrete significant absorption feature required
by this component (and not modeled with only two
absorbers). Therefore, we consider the detection of this component as
tentative. The
absorbing components (whether two or three different absorbers are
present) are in pressure balance (within $1\sigma$) and have outflow
velocities indistinguishable from each other.  Thus, our results give
further support to the idea that ionized absorbers in AGN form a
multi-phase medium, where a hot gas component is confining the
other(s). However,  in the model with only two absorbers, one is
located in an unstable branch of this curve. On the other hand,  
in the three-phase model all the components lie in
stable regions of the thermal equilibrium curve. Thus, only the model with three absorbing
components allows a multi-phase absorber to survive for long periods
of time. In this case, the multi-phase medium would not dissolve even
if the crossing time is much larger than the timescale in which the
medium adapts to mechanical perturbations.

\acknowledgements 
We thank the referee for constructive comments that helped to improve
the paper. 
This research is based on observations obtained with XMM-Newton, an
ESA science mission with instruments and contributions directly funded
by ESA Member States and NASA. This work was supported by the UNAM
PAPIIT grant IN118905 and the CONACyT grant J-49594. EJB and MSLl
have been supported by Spanish MEC under grant AYA2001-3939-C03-02.
YK and EJB aknowledge support by the ESA faculty program.


\clearpage

\begin{deluxetable}{llllllllcccc}
\tablecolumns{12} \tablewidth{0pt}  \rotate \tablecaption{Continuum
  parameters and goodness of the tested models \label{tab:models}}
\tabletypesize{\scriptsize} \tablehead{\colhead{\bf M} &
  \multicolumn{2}{c}{\bf Power Law} &\multicolumn{3}{c}{\bf Iron Line}
  &\multicolumn{2}{c}{\bf Black Body}  &  \multicolumn{3}{c}{\bf WA} &
  \colhead{\bf Goodness} \\ & \colhead{$\Gamma$}  & \colhead{A$_{\rm
      pwlw}$\tablenotemark{a}} & \colhead{E (keV)}  &
  \colhead{$\sigma$ (keV)} & \colhead{A$_{\rm Line}$\tablenotemark{b}}
  & \colhead{kT (keV)} & \colhead{A$_{\rm BB}$\tablenotemark{c}} &
  \colhead{SH} &  \colhead{H}&  \colhead{L} &  \colhead{$\chi^2$} }
\startdata 
A &  1.416$\pm$0.015 & 1.63$\pm0.03$ &-- &-- &-- &-- &-- &
--& -- & --& 231 for 174 dof\\ 
B &  1.450$\pm$0.016 & 1.69$\pm$0.04 &
6.421$\pm$0.025 & 0.14$\pm$0.03 & 1.86$^{+0.30}_{-0.28}$ & --  &-- &--
&-- & --& 161 for 171 dof\\ 
C & 1.20$\pm$0.10    & 2.33$\pm$0.11 & --
& -- & -- & 0.07$\pm$0.03 & 1.31$\pm$0.06 & --      & --      &
$\surd$ & 475 for 410 dof \\ 
D & 1.71$\pm$0.10    & 3.25$\pm$0.25 &
--  & -- & -- & 0.08$\pm$0.02 & 1.52$\pm$0.15 & --      & $\surd$ &
$\surd$ & 457 for 407 dof\\
E & 1.73$\pm$0.12    & 3.59$\pm$0.19 & 
--  & -- & -- & 0.08$\pm$0.02 & 1.72$\pm$0.17 & $\surd$ & $\surd$ &  
$\surd$ & 439 for 404 dof\\
F & 1.572$^{+0.018}_{-0.012}$ & 2.89$^{+0.08}_{-0.06}$ (rgs) &
$6.42\pm0.025$ & 0.14$\pm0.03$ & -- & 0.09$\pm{0.01}$ &
1.06$\pm0.08$ (rgs)& $\surd$ & $\surd$ & $\surd$ &  810 for
648 dof\\
 & &  $ 1.98\pm{0.09}$(pn) &  & & 1.8$\pm0.2$ (pn) & &
$0.60\pm{0.02}$ (pn) \\
\enddata 
\tablenotetext{a}{In units of $10^{-3}$ ph s$^{-1}$ cm$^{-2}$
  keV$^{-1}$ at 1 keV} \tablenotetext{b}{Flux in units of $10^{-5}$ ph
  s$^{-1}$ cm$^{-2}$} \tablenotetext{c}{In units of $10^{-4}$
  $L_{39}/D_{10}^2$, where $L_{39}$ is the source luminosity in units
  of $10^{39}$ erg s$^{-1}$  and $D_{10}$ is the distance to the
  source in units of 10 kpc}
\end{deluxetable}


\clearpage
\begin{deluxetable}{llllllllll}
\tablecolumns{10} \tablewidth{0pt}  \rotate \tablecaption{Ionized
  absorber parameters of the tested models \label{tab:ioniz} }
\tabletypesize{\footnotesize}
\tablehead{\colhead{\bf Model} & \multicolumn{3}{c}{\bf Low-Ionization
    PHASE} &  \multicolumn{3}{c}{\bf High-Ionization PHASE} &
  \multicolumn{3}{c}{\bf Super High-Ionization PHASE} \\ &
  \colhead{$\log {U}$} & \colhead{$\log {N_H}$} & \colhead{$V_{\rm
      out}$} & \colhead{$\log {U}$} & \colhead{$\log {N_H}$} &
  \colhead{$V_{\rm out}$} & \colhead{$\log {U}$} & \colhead{$\log
    {N_H}$} & \colhead{$V_{\rm out}$}}  \startdata 
C & & & & $1.26^{+0.06}_{-0.07}$ & $21.60\pm{0.12}$ & $106\pm{204}$  \\ 
D & $-0.06^{+0.12}_{-0.13}$ & $21.11\pm{0.09}$ & $267\pm{159}$  & $1.31\pm{0.12}$ & $21.99\pm{0.11}$ & $272\pm{149}$  \\
E & $-0.04^{+0.30}_{-0.09}$ & $21.08\pm{0.08}$ & $273\pm{134}$  & $1.09\pm{0.07}$ & $21.61\pm{0.15}$ & $296\pm{138}$ & $1.75\pm{0.07}$ & $22.01\pm{0.18}$ & $270\pm{168}$ \\
F & $0.05^{+0.30}_{-0.16}$ & $21.08\pm{0.18}$ & $273\pm{134}^{a}$ & $1.05\pm{0.06}$ & $21.55\pm{0.41}$ & $296\pm{138}^{a}$ & $1.72\pm{0.09}$ & $21.71\pm{0.21}$ & $270\pm{168}^{a}$ \\ 
\enddata
\tablenotetext{a}{Velocities constrained to the best-fit value
  obtained for model E}
\end{deluxetable}

\clearpage
\begin{deluxetable}{lccl}
\tablecolumns{4} \tablewidth{0pc} \tablecaption{Parameters of the
  emission lines \label{tab:lines}} \tablehead{\colhead{\bf Line} &
    \colhead{\bf Rest Wavelength$^{a}$} & \colhead{\bf  Width} &
  \colhead{\bf  Flux$^{b}$} \\  & \colhead{\AA} &
  \colhead{km s$^{-1}$}}
\startdata 
OVIII (K$_\alpha$) & 18.97     & 650$\pm820$ & 4.9$^{+4.1}_{-1.7}$      \\ 
OVII (w)           &21.60     & 620$\pm390$ & 3.1$^{+1.7}_{-1.3}$ \\ 
OVII (x+y)         & 21.85    & 740$\pm870$ & 3.6$^{+1.4}_{-1.2}$ \\ 
OVII (z) & 22.107$\pm{0.010}$ & 730$\pm570$ & 5.9$\pm{1.0}$      \\ 
\hline
\enddata 
\tablenotetext{a}{ The position of the
emission lines was constrained to have the same redshift found for the
OVII (z) line, which was consistent with the rest frame of NGC~985.}
\tablenotetext{b}{Intrinsic flux in
  units of $10^{-14}$ erg s$^{-1}$ cm$^{-2}$.}
\end{deluxetable}

\clearpage

\clearpage
\clearpage


\clearpage

\clearpage

\clearpage

\clearpage

\clearpage

\end{document}